**Large Second Harmonic Kerr rotation in GaFeO$_3$ thin films on YSZ buffered Silicon.**

Darshan C. Kundaliya[*], S. B. Ogale, S. Dhar
*Center for Superconductivity Research, Department of Physics, University of Maryland, College Park, MD 20742*

K. F. McDonald, E. Knoesel, T. Osedach, S. E. Lofland,
*Department of Physics and Astronomy, Rowan University, Glassboro, New Jersey 08028-1701*

S. R. Shinde and T. Venkatesan
*Center for Superconductivity Research, Department of Physics, University of Maryland, College Park, MD 20742*

**Abstract**

Epitaxial thin films of gallium iron oxide (GaFeO$_3$) are grown on (001) silicon by pulsed laser deposition (PLD) using yttrium-stabilized zirconia (YSZ) buffer layer. The crystalline template buffer layer is *in-situ* PLD grown through the step of high temperature stripping of the intrinsic silicon surface oxide. The X-ray diffraction pattern shows *c*-axis orientation of YSZ and *b*-axis orientation of GaFeO$_3$ on Si (100) substrate. The ferromagnetic transition temperature ($T_C \sim 215$ K) is in good agreement with the bulk data. The films show a large nonlinear second harmonic Kerr rotation of ~15 degrees in the ferromagnetic state.

*Corresponding author:  Tel:* +1-301-405-7672   *Fax:* +1-301-405-3779
*Email address:* darshan@squid.umd.edu

## 1. Introduction

Multifunctional materials have attracted significant interest in recent years due to the coupling of their ferromagnetic and piezoelectric/ferroelectric properties which is of great interest from the fundamental as well as applied points of view [1-5]. GaFeO$_3$ (GFO) is an interesting example in this context which exhibits ferromagnetic and pyroelectric properties simultaneously [6]. This material was first discovered by Remeika et al.[6] and its crystallographic properties (orthorhombic structure) were described by Abrahams et al. [7] on the basis of unit cell volume and space group given by Wood et al. [8]. Observation and possible mechanisms of magnetoelectric effects in this ferromagnet were described by G.T. Rado [9] and Frankel et al. [10], and a sizable magnetic anisotropy was reported in Ga$_{2-x}$Fe$_x$O$_3$ single crystals by Levine et al. [11].

Recently, magnetization-induced second harmonic generation (MSHG) and X-ray directional dichroism have been investigated in single crystal GFO [12,13]. Remarkably large effects have been found due to the intrinsically non-centro-symmetric nature of this crystallographic system. For building technologically viable device systems it is important to develop such multifunctional materials as crystalline thin films on different substrates, most desirably on Si. In this paper we describe the growth, observation of large SHG Kerr effect, and the magnetic properties of GFO thin films grown by pulsed-laser deposition on crystalline YSZ-template-buffered Si. We also present results for the films grown on crystalline YSZ substrates.

## 2. Experimental

High purity powders of Ga$_2$O$_3$ (99.999%) and Fe$_2$O$_3$ (99.998%) were ground in the agate-mortar in the proper stoichiometric ratio. The resultant mixture were cold pressed in the pellet form and calcined at 1050ºC for 24 hrs. These pellets were reground thoroughly, pelletized and sintered at 1150ºC for 24 hrs to obtain stoichiometric GFO. High-purity commercially available YSZ target was used for the corresponding buffer layer growth on (001) Si. A KrF excimer laser ($\lambda$=248 nm) was used for ablation. The laser energy was held at 1.8-2.0 J/cm$^2$. The base pressure was below $1\times10^{-6}$ Torr before every deposition. High-quality epitaxial growths of GFO on YSZ substrate (100) and on YSZ buffered Si (100) were realized at an oxygen growth pressure of 400 mTorr. After each deposition the sample was cooled to room temperature in the same oxygen pressure of 400 mTorr. The substrate temperature was kept at 650 ºC for GFO deposition. Growth of crystalline template buffer layer of YSZ on (001) Si was realized without the need for any chemical process for removal of the surface oxygen as described in ref. 14. This method avoids the use of highly toxic HF solution to strip the oxide from the surface of Si substrate.

Film thickness and epitaxial quality were analyzed by Rutherford backscattering ion-channeling spectroscopy. Magnetization measurements were performed with a SQUID magnetometer (Quantum Design, MPMS). Second-harmonic generation studies were carried out with a Coherent Ti-sapphire femtosecond laser tuned to 750 nm. The incoming *s*-polarized light was nearly normal to the film surface and was focused to a spot size of about 100 μm. A photomultiplier tube was used to detect the second harmonic light from the film. SHG measurements were done at room temperature (above $T_C$) and at low temperature (100 K, i.e. below $T_C$) in an applied field of 3 kOe, which was applied in the film plane, normal to the polarization.

## 3. Results and Discussion

GFO crystallizes into orthorhombic structure with lattice constants $a = 8.72$ Å, $b = 9.37$ Å and $c = 5.07$ Å. Lattice constants of YSZ and Si substrates are 5.12 Å and 5.41 Å respectively. The lattice matching between the three materials is as under: GFO/YSZ is 99% matched along c-plane and YSZ/Si is 94.7% matched. Figure 1 (a) and (b) show the X-ray diffraction (XRD) patterns of GFO grown on YSZ substrate (100) and on YSZ buffered Si (100) respectively showing b-axis orientation of the GFO layer. The high crystalline quality of the YSZ buffer layer is also clear. The b-axis growth of the GFO layer is significant since it yields the highest nonlinear Kerr effect for this crystal[12]. Figure 1(c) shows the rocking curve full width at half maximum (FWHM) of GFO/YSZ and GFO/YSZ/Si. Further, the FWHM of GFO/YSZ is ~0.44 ° (as compared to 0.24° of single crystal YSZ substrate), establishing a high orientational quality of the film.

The ion backscattering channeling data for GFO on (001) single crystal YSZ is shown in Fig. 2. The random spectrum matches well with the simulated one. The significant reduction in the backscattering yield for the Ga signal (minimum yield $\chi_{min} \cong 12\%$) upon channeling confirms the XRD observations of high quality epitaxial growth of the films on single-crystal YSZ substrate. The commercial YSZ single crystalline substrate used for growth showed $\chi_{min} \cong 10\%$, suggesting that the observed 12% channeling in GFO/YSZ reflects excellent film quality. The ion channeling of YSZ buffer on Si (thickness=1250 Å) gives $\chi_{min} \cong 8.5\%$ (Fig. 2(b)), suggesting a very good epitaxial quality of YSZ template on Si for the growth of GFO. However, when GFO was grown on this high quality YSZ/Si, the corresponding orientational registry as observed by RBS ion channeling was found to be somewhat inferior ($\chi_{min} \cong 45\text{-}50\%$) to that for film directly grown on crystalline YSZ. It may be noted that a slight tilt of crystallites from the vertical increases the $\chi_{min}$ dramatically. Indeed, we noted an increase in rocking curve FWHM to ~0.70º for GaFeO$_3$ on YSZ/Si (Fig. 1(c)). Nonetheless, it is important to note that we do not see any extra peaks in θ-2θ XRD scans and the highly *b*-axis orientation persists in the GFO thin films. The reasons for the observed features could be traced to the lattice and thermal stress effects, and their evolution during growth. Because of the large in-plane tensile strain of GFO with silicon vis a vis that with pure YSZ crystal, the film of silicon would release stress more readily by production of extended defects and the properties of stress free regions of the film should be close to that of bulk GFO materials, as against those of films on YSZ crystal. This is indeed observed as discussed later.

We also performed magnetic measurements on GFO/YSZ and GFO/YSZ/Si thin films and on bulk sintered GFO, the results of which are shown in Fig. 3. Figure 3 (a) shows the plots of magnetization *M* vs applied magnetic field *H* (*H* ∥ *c* axis) for GFO/YSZ and GFO/YSZ/Si, respectively, measured at 5 K. As stated earlier, GFO is magnetically highly anisotropic with spontaneous saturation magnetic moment 0.79 $\mu_B$/Fe along the *c* axis (magnetic easy axis) for GFO/YSZ. Figure 3 (b) is the plots of the *M* vs. *T* in an applied field of 50 Oe (*H* ∥ *c* axis) for GFO bulk, GFO/YSZ and GFO/YSZ/Si. $T_C$ values are ~200 K and ~215K for the two cases, respectively, and are in very good agreement with the bulk result. Interestingly, the behavior of *M* vs. *T* is nearly Brillouin-like.

We also observed a large nonlinear second harmonic magneto-optical Kerr effect in our GFO/YSZ/Si thin films. Figure 4 (a) and (b) compare the analyzer angle (θ) dependence of the second harmonic light at room temperature (above $T_C$) and at 100K (below $T_C$). At room temperature SH light is relatively small, principally *p*-polarized and independent of field. As the temperature is lowered to 100 K ($< T_C$), the intensity of the light increases, and the polarization of SH light rotates with respect to the direction of magnetic field by ~ 15º away from *p* polarization. In bulk single crystal, Ogawa et al. [12] have observed SH Kerr rotation of nearly 73º at 100 K. This result cannot be directly compared with that of a thin film since the film has additional symmetry influencing contributions due strain effects giving rise to a thickness dependence of the SHG signal in this non-centro-symmetric system and also the contributions of surfaces/interfaces/extended defects. While there may be room for further improvement in the strength of the effect, the magnitude of the effect already observed is quite significant for potential applications. Interestingly, we did not observe strong SHG in GFO grown directly on YSZ probably because of strain induced transformation of the film to nearly centro-symmetric configuration. As stated earlier, in case of the more strained GFO/YSZ/Si configuration, the release of stress through creation of extended defects should lead to bulk GFO like properties with non-centro-symmetry effects, as indeed observed.

In conclusion, we have systematically studied the growth, magnetization and nonlinear second harmonic Kerr effect of $GaFeO_3$ thin films on single crystal (001) YSZ and YSZ-buffered Si substrates. The magnetic transition temperature $T_C$ is ~215 K. A large nonlinear SH Kerr rotation is observed at 100 K with a rotation angle of 15º.

We acknowledge support under MRSEC grant DMR 00-80008.

**Figure Caption**

Figure 1	X-ray diffraction θ-2θ scans for the thin films of (a) GFO/YSZ (b) GFO/YSZ/Si. (c) Rocking curve FWHM of GFO/YSZ (bottom) and GFO/YSZ/Si (top) suggesting good epitaxial quality of the films. The asterisk sign indicates an appearance of small $SiO_x$ phase.

Figure 2	1.5 MeV $He^+$ ion RBS spectra in random and aligned directions for GFO films grown on YSZ. Minimum Yield obtained for GFO/YSZ is 12% and for YSZ in YSZ/Si is 7.5% suggesting good epitaxial relation of GFO and YSZ on the respective substrates.

Figure 3	Magnetization isotherms (*M-H*) of (a) GFO/YSZ and GFO/YSZ/Si give a saturation moment of 0.79 $\mu_B$/Fe and 0.39 $\mu_B$/Fe respectively. (b) temperature (*T*) dependent magnetization (*M*) plots of GFO bulk, GFO/YSZ and GFO/YSZ/Si thin films.

Figure 4	Analyzer angle (θ) dependence of the second harmonic light (a) at room temperature (above $T_C$) and (b) at 100 K (below $T_C$). θ = 0 corresponds to *p* polarization and open (closed) circles correspond to a negative (positive) field of 3 kOe.

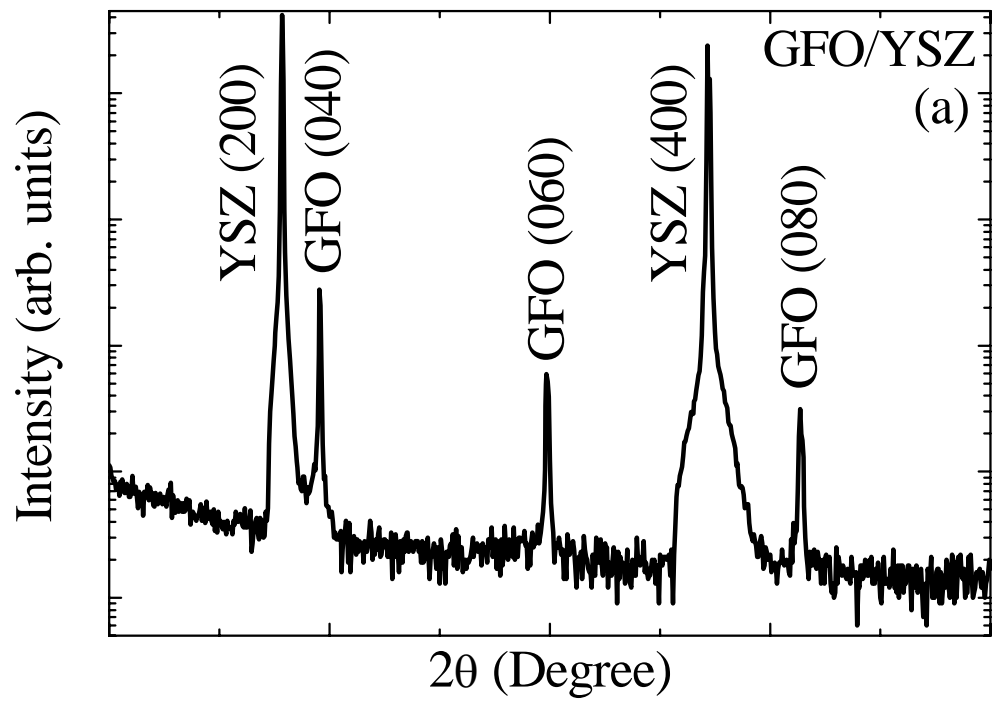

Figure 1(a) *Kundaliya et al.*

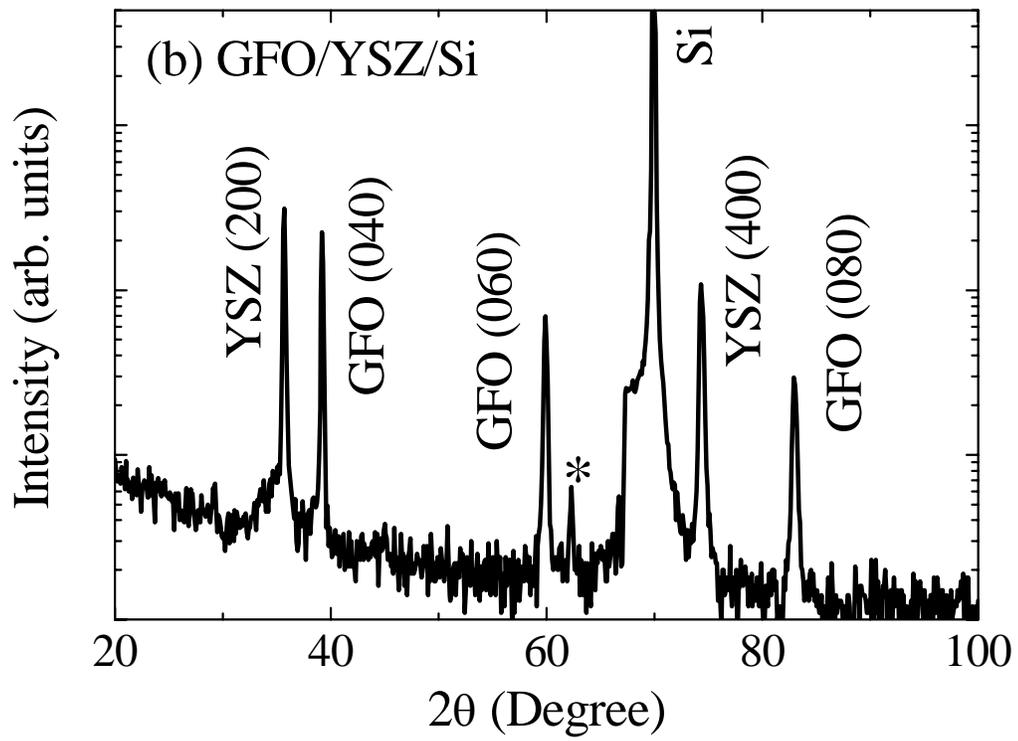

Figure 1(b) Kundaliya *et al.*

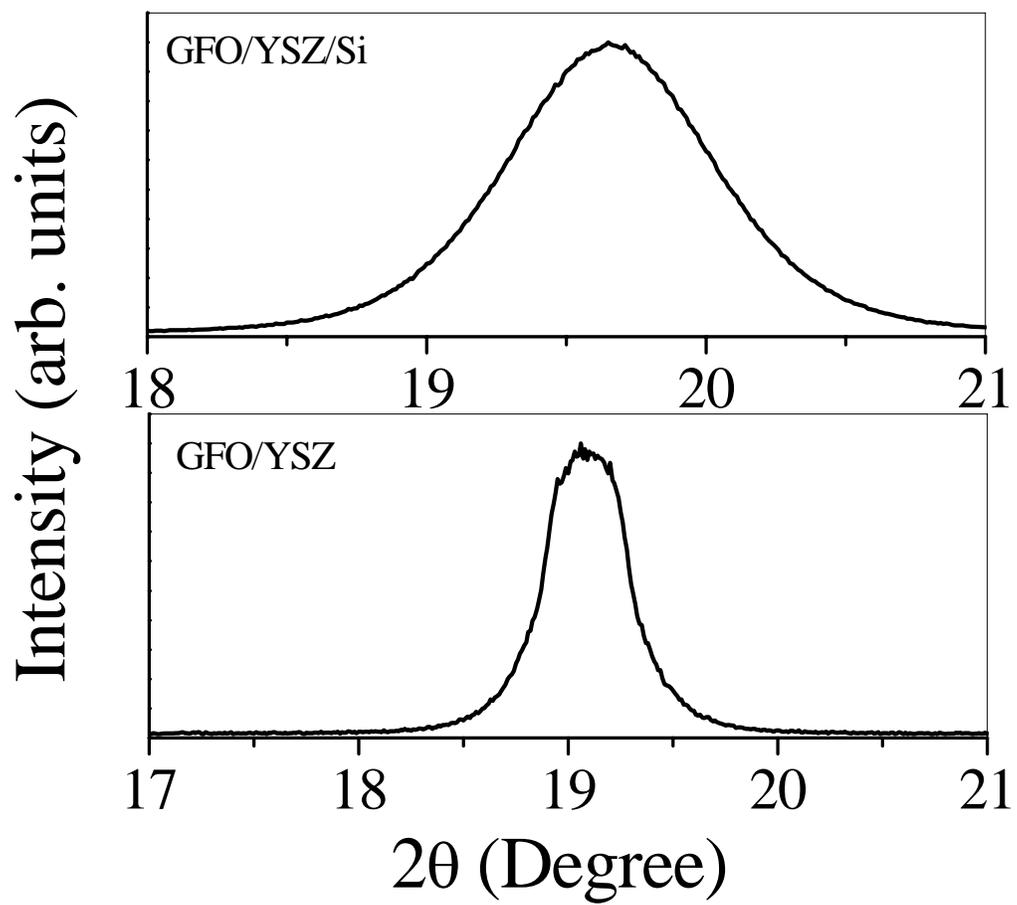

Figure 1(c) Kundaliya *et al.*

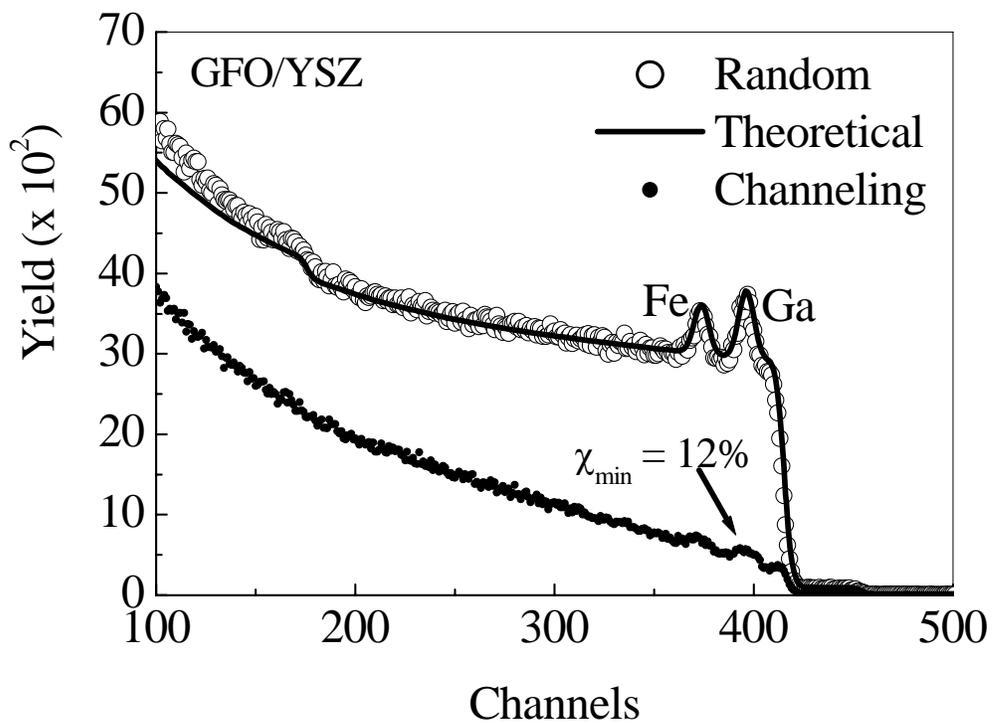

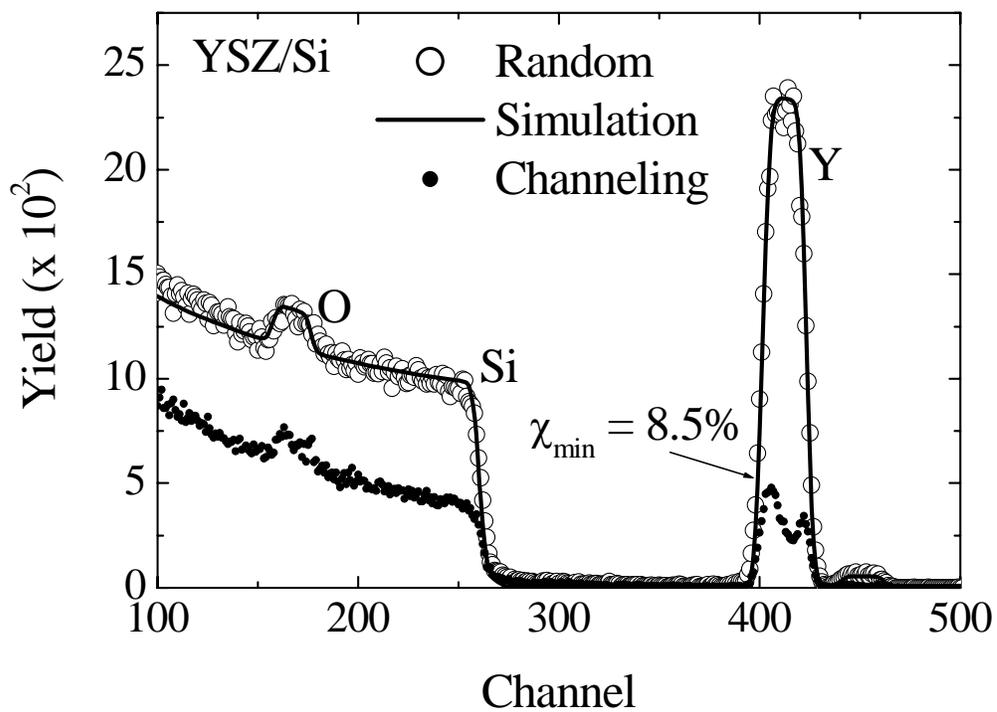

Figure 2 Kundaliya *et al.*

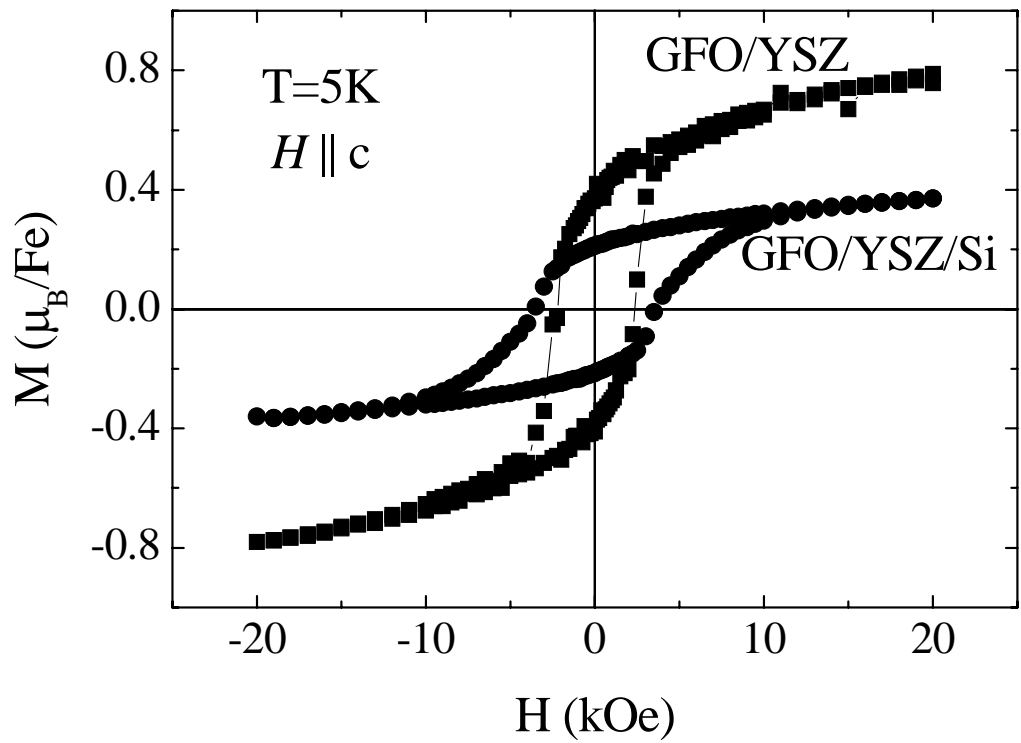

Figure 3(a) Kundaliya *et al.*

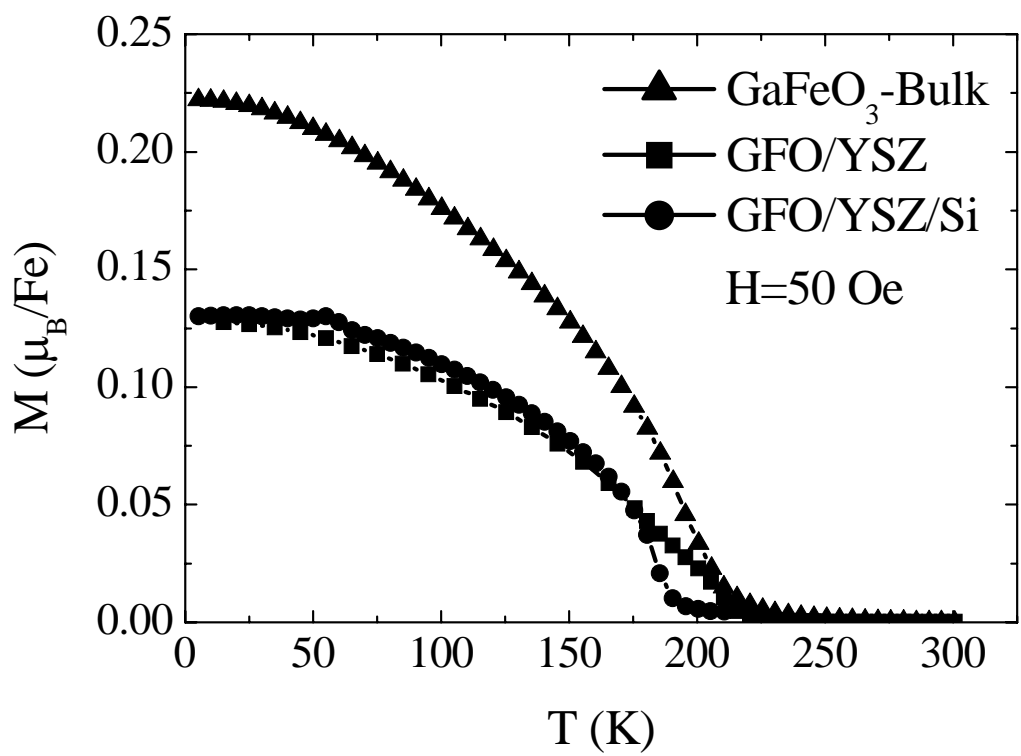

Figure 3(b) Kundaliya *et al.*

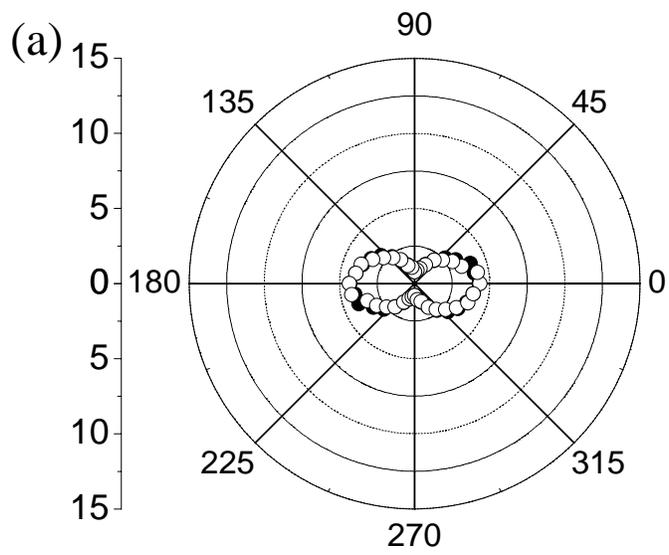

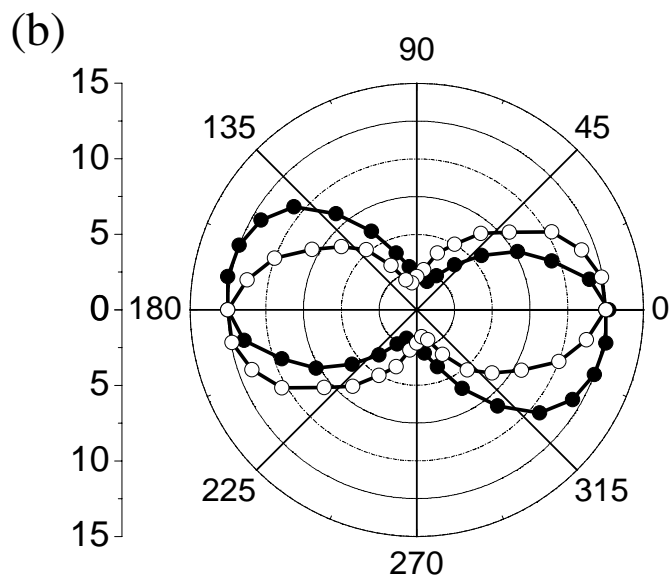

Figure 4(a) & (b) Kundaliya *et al.*